%% file: main.tex
\documentclass[lettersize,journal]{IEEEtran}
\usepackage{amsmath,amsfonts}
\usepackage{algorithmic}
\usepackage{algorithm}
\usepackage{array}
\usepackage[caption=false,font=normalsize,labelfont=sf,textfont=sf]{subfig}
\usepackage{textcomp}
\usepackage{stfloats}
\usepackage{siunitx}
\usepackage{url}
\usepackage{verbatim}
\usepackage{graphicx}
\usepackage{orcidlink}
\usepackage{cite}
\usepackage{import}
\DeclareSIUnit\sample{Sa}

\begin{document}

\title{A Modular Zero-Dead-Time Data Acquisition and Real-Time GPU Processing Platform for High Throughput Physics Experiments}

\author{Toma-Stefan Cezar\orcidlink{0009-0009-5121-1821}, 
        Marios Maroudas\orcidlink{0000-0003-1294-1433}, 
        and Dieter Horns\orcidlink{0000-0003-1945-0119}%
\thanks{The authors are affiliated with the Institut für Experimentalphysik, Universität Hamburg, Luruper Chaussee 149, D-22761 Hamburg, Germany (e-mail: toma-stefan.cezar@studium.uni-hamburg.de).}}

\markboth{Prepared for submission to IEEE Transactions on Instrumentation and Measurement}%
{Cezar \MakeLowercase{\textit{et al.}}: A Modular Zero Dead Time Data Acquisition Platform}

\maketitle

\begin{abstract}

\textit{High-throughput physics experiments require efficient and increasingly complex real-time processing. This paper presents a modular, software-defined platform combining high-bandwidth PCIe digitizers with consumer GPUs to achieve continuous, zero-dead-time data acquisition. Utilizing NVIDIA CUDA, the system provides a scalable pipeline for real-time fast Fourier transforms and statistical averaging. Benchmarks demonstrate that the platform can sustain continuous processing at sampling rates up to \SI[per-mode=symbol]{500}{\mega\sample\per\second}, effectively managing data throughputs of \SI[per-mode=symbol]{1}{\giga\byte\per\second}. To validate the \textit{in-situ} zero-dead-time architecture, end-to-end phase continuity tests were conducted, constraining fractional data loss to below $10^{-12}$. Furthermore, long-term system stability was demonstrated through an uninterrupted one-month data acquisition run. In its current deployment for the WISPLC dark matter experiment, the platform operates at \SI[per-mode=symbol]{124}{\mega\sample\per\second} with a resolution bandwidth of \SI{0.1}{\hertz}. This implementation enabled a significant reduction in data storage requirements using real-time spectral averaging. The callback-driven software architecture, multi-GPU workload distribution, and custom hardware shielding solutions are detailed, establishing this platform as a flexible and cost-effective alternative to traditional hardware-based pipelines.}

\end{abstract}


\section{Introduction}
\label{sec:introduction}

\IEEEPARstart{M}{odern} high-throughput physics experiments rely on robust data acquisition (DAQ) pipelines to digitize and process high-bandwidth analog signals. A standard DAQ system uses an analog-to-digital converter (ADC) to convert detector outputs into digital data. In conventional setups, subsequent data processing occurs offline. However, the escalating data volumes in current experiments necessitate real-time processing to reduce storage requirements and enable flexible transient signal searches.

Hardware-based real-time processing systems, particularly those utilizing Field Programmable Gate Arrays (FPGAs) or Radio Frequency Systems on a Chip (RFSoCs), are the current industry standard for high-speed data acquisition and signal processing \cite{Stefanazzi_2021_qick, HADES_2011_trigger}. These systems offer exceptional computational efficiency and deterministic latency, which is critical for experiments requiring nanosecond precision. However, they present significant development bottlenecks. Developing and modifying FPGA-based firmware requires specialized hardware description language (HDL) expertise, such as VHDL or Verilog \cite{edwards_2007_challenges}. Furthermore, the synthesis and routing processes inherent to FPGAs lead to long compilation times and complex debugging cycles \cite{xiao_2019_fpga_compile}. While RFSoCs provide high integration by combining ADCs and FPGAs on a single chip, they remain expensive, specialized devices with a high entry barrier for general laboratory use \cite{javaid_2022_performance, Murthy_2025_comparative}. Achieving optimal performance on these platforms requires the precise configuration of numerous interdependent parameters, such as decimation factors and clock frequencies, which increases system complexity. Consequently, these architectures can become rigid and difficult to maintain in experimental environments where processing algorithms require frequent iteration.

As an alternative, this paper presents a modular, software-defined DAQ platform that combines high-bandwidth PCIe-based ADC cards with standard consumer electronics. Software-based processing has been explored for axion searches, such as in the CULTASK \cite{Lee_2017} and CAPP \cite{Ahn_2022} experiments, but these CPU-based implementations remain limited in throughput and resolution bandwidth. The utility of a GPU-accelerated readout was previously demonstrated in the WISPDMX experiment \cite{Nguyen_2019_wispdmx}. Our platform builds upon these concepts by offloading real-time data processing to NVIDIA CUDA-capable graphics processing units (GPUs) to achieve high computational throughput while utilizing standard C and C++ programming. This approach reduces setup costs and development time compared to traditional FPGA pipelines. The main contributions of this work are as follows:

\begin{itemize}
    \item Realization of a modular, software-defined data acquisition architecture that enables continuous high-throughput signal acquisition and real-time processing using PCIe-based digitizers and GPU-based processing.

    \item Implementation of a callback-driven design to enable flexible real-time computation of derived signal statistics such as power, variances, and spectral averages.

    \item \textit{In-situ} validation of zero dead time capabilities through the demonstration of phase consistency over extended measurement periods.

    \item Deployment of multi-layered software abstractions to ensure compatibility with diverse data acquisition hardware platforms.

\end{itemize}

The platform is currently deployed in the WISPLC dark matter experiment at the University of Hamburg \cite{WISPLC}. It also provides a flexible, software-defined alternative with a larger dynamical range in comparison with specialized digital base band converter (DBBC) systems \cite{Tuccari_2012_dbbc} utilized in broadband searches such as BRASS \cite{Brass}. By implementing software-defined logic rather than fixed hardware architectures, our platform supports rapid iteration of diverse signal processing operations without requiring specialized firmware development. The software enables real-time fast Fourier transforms (FFTs) with highly configurable resolution bandwidths (RBW) and continuous statistical averaging. Additional digital signal processing and filtering, as well as condition-based data storage can be implemented. By leveraging real-time processing and averaging, the raw data storage requirements of the WISPLC experiment were reduced by orders of magnitude. A detailed description of the hardware constraints, multi-threaded software architecture, and performance benchmarks is provided, while the underlying C/C++ and CUDA codebase will be open-sourced.

\section{The Platform Architecture}
\label{sec:architecture}

The proposed data acquisition platform consists of custom C/C++ and CUDA software integrated with a hybrid hardware setup. This setup merges specialized scientific digitizers with standard consumer electronics. The only required external hardware is a suitable amplification stage. The following sections detail the hardware constraints, the concurrent software architecture, and the implementation of user-defined processing callbacks.

\subsection{Hardware requirements}
\label{sec:hardware}

The core system requires a high-performance PCIe analog-to-digital converter (ADC) paired with a modern multi-threaded central processing unit (CPU) and one or more CUDA-capable graphics processing units (GPUs). In addition to the host computing architecture, the analog signal chain requires a suitable amplification stage and a suitable low-pass filter for anti-aliasing prior to digitization to prevent high-frequency foldover.

The platform architecture is hardware agnostic, utilizing a modular interface that can support any PCIe ADC card employing a first-in-first-out (FIFO) ring buffer structure. For our specific experimental deployment, we utilize Spectrum Instrumentation SPCM series digitizers, specifically the M4i.4420-x8 \cite{Spectrum_M4i_Manual} and M2p.5941-x8 models \cite{Spectrum_M2p_Manual}. These cards provide the high data throughput and bit resolution necessary for the real-time spectral analysis required by the WISPLC experiment \cite{WISPLC}. The M4i.4420-x8 provides continuous sampling rates up to \SI[per-mode=symbol]{250}{\mega\sample\per\second} across two channels with 16-bit resolution and a programmable bipolar input range between $\pm\SI{200}{\milli\volt}$ and $\pm\SI{10}{\volt}$. In most experimental configurations, this ADC, combined with a high-stability external reference clock, offers adequate performance. Integrating a different ADC vendor requires only minimal modifications to the abstract C backend to interface the specific hardware drivers.

The host system must support multiple PCIe devices at full bandwidth to prevent bus bottlenecks between the ADC and GPUs. The GPUs execute the real-time processing pipeline. Although computational performance, such as floating-point operations per second, and memory bandwidth determine the achievable processing rate, video random-access memory (VRAM) capacity typically imposes the most significant limitation. To overcome this limitation, the software is designed to distribute workloads across multiple GPUs. While a single GPU with sufficient VRAM is the ideal configuration, our specific deployment utilizes a triple GPU setup to achieve the required total memory using available consumer hardware. This multi-GPU approach enables higher resolution Fourier transforms with narrower resolution bandwidths than the single-GPU setup, albeit at the cost of reduced scaling efficiency due to non-ideal VRAM distribution across devices, as detailed in Section \ref{sec:callbacks}.

\begin{figure}
\centering
\includegraphics[width=\linewidth]{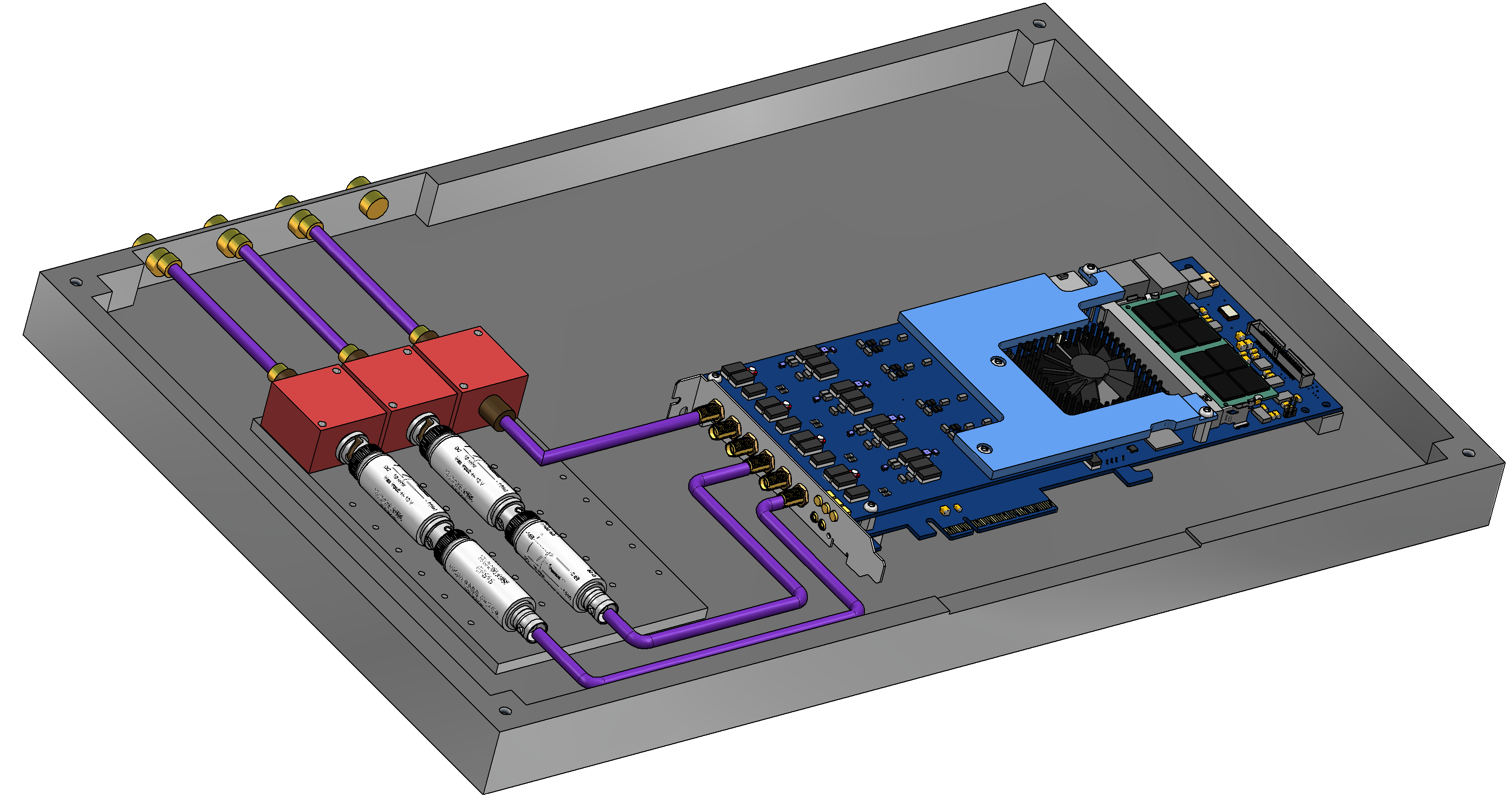}
\caption{Design of the custom aluminum Faraday cage utilized to isolate the ADC from host system electromagnetic interference.}
\label{fig:faraday_cage}
\end{figure}

Finally, to mitigate electromagnetic interference (EMI), the ADC should be physically isolated from the host PC components using a PCIe riser cable. Enclosing the ADC card and secondary RF components within a custom-made grounded Faraday cage, constructed from \SI{10}{\milli\meter}-thick aluminum, effectively eliminates high-frequency noise. In particular, this external shielding is necessary due to non-ideal shielding along the analog input trace of the ADC, which is susceptible to RF interference from nearby high-power digital components such as the GPU and switching power supplies. Because enclosing an active digitizer restricts ambient airflow, passive heatsinks and active ventilation fans were integrated directly into the custom enclosure. This cooling solution guarantees thermal stability, preventing temperature-induced baseline drift during extended acquisition runs. The physical design of this enclosure is shown in Fig.~\ref{fig:faraday_cage}.

\subsection{General Software structure}
\label{sec:software}

The current software is developed using a mix of C and C++\footnote{We rely on C++ due to external dependencies like CUDA and Dear ImGUI \cite{dearimgui}, but the codebase strictly follows C-style programming conventions.}. It includes additional CUDA modules for GPU-based data processing. Although the implementation is primarily tailored to Spectrum Instrumentation ADC cards, it contains an abstract backend interface. This interface enables support for other ADC devices with minimal modification.

\begin{figure}
\centering
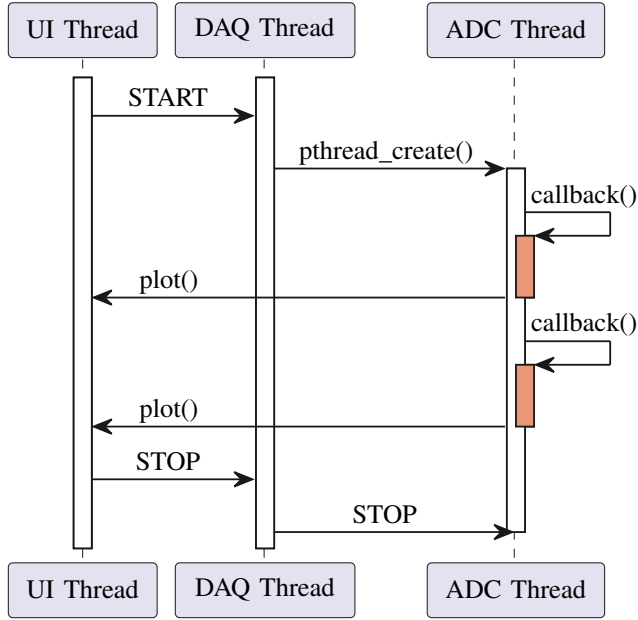
\caption{Sequence diagram illustrating the concurrent thread architecture. The UI thread issues execution commands to the DAQ thread, which subsequently spawns the ADC worker thread via \texttt{pthread\_create()}. The ADC thread handles the hardware ring buffer, triggers the user-defined data processing callbacks, and issues \texttt{plot()} commands back to the UI thread for real-time visualization without blocking data acquisition.}
\label{fig:thread_structure}
\end{figure}

The software architecture is subdivided into multiple concurrent threads utilizing the POSIX Threads (pthreads) API \cite{ieee_posix} to ensure high performance and low latency. The thread architecture is illustrated in Fig.~\ref{fig:thread_structure}. By assigning specific tasks, such as data acquisition, GPU memory transfer, and disk I/O, to independent threads, the platform can maintain a continuous, zero-dead-time pipeline even at maximum sampling rates. The system currently provides two long-lived primary threads: a DAQ control thread and a user interface (UI) thread. The DAQ control thread processes execution commands issued by either the main thread or the UI thread. Upon receiving a request, the DAQ control thread executes the corresponding operation, which includes spawning a dedicated DAQ worker thread when data acquisition is initiated. This worker thread initializes the ADC card and the required data transfer ring buffer. After initialization, the worker thread starts the data acquisition by enabling the ADC card. By utilizing the hardware FIFO mode (see Sect.~\ref{sec:hardware}), the system continuously acquires data without any dead time. The ADC card continuously writes into a circular ring buffer in host memory. Once it reaches the end of the allocated memory, it wraps around to the beginning.

During this continuous process, the ADC driver periodically updates the worker thread with the volume of newly available data. Once this available data exceeds a user-defined threshold, the thread invokes a user-provided callback function. This callback receives a pointer to the beginning of the valid data segment and its length. Through this abstraction, the user can treat the continuous data stream as a discrete linear buffer of a fixed size. After the callback function finishes execution, the processed data segment is marked as processed. This allows the ADC hardware to safely overwrite it. Standard ring buffer implementations handle wraparounds by copying fragmented data into an internal linear buffer if a valid chunk crosses the memory boundary. Because this memory transfer is computationally expensive, our architecture avoids it entirely. We enforce a ring buffer size that is a strict integer multiple of the user-defined threshold. This guarantees that every data chunk aligns perfectly with the buffer boundaries, preventing fragmented reads.

 As long as the execution time of the callback is lower than the time required for the ADC to fill the threshold, and the buffer is large enough to absorb operating system scheduling jitter, this approach ensures true zero dead time (see Section~\ref{sec:benchmarks}). This abstraction is highly advantageous. The user only needs to write a processing callback without managing the underlying hardware or memory wrapping logic. Furthermore, this structure enables straightforward future expansion to additional features. Examples include supporting multiple ADC cards, processing data streamed through network interface cards (NICs), or integrating digital-to-analog converters (DACs) for signal synthesis.

\subsection{User defined Callbacks}
\label{sec:callbacks}

The user-defined callback function implements the real-time data processing. We present our implementation for computing real-time FFTs on a single-GPU or triple-GPU setup. This system is highly modular. By modifying the callback function, the pipeline can be adapted for different use cases, allowing users to implement custom digital filtering, real-time downconversion, or specialized transient signal detection algorithms. We provide general-purpose functions for FFTs, thread-safe memory buffer handling, downsampling, plotting, and HDF5 file handling \cite{hdf5}.

\subsubsection{Baseline Single-GPU Setup}

Using a single GPU provides the advantage of continuous VRAM memory. This allows for data processing without inter-GPU transfers. The downside of this configuration is the limited VRAM, which restricts the achievable maximum resolution bandwidth (RBW). For simplicity, we execute one continuous FFT per input channel utilizing NVIDIA's \texttt{cuFFT} library \cite{cuFFT}. The RBW of a single FFT depends entirely on the duration of data acquisition. For a specific sample rate $f_s$, the DAQ time is dictated by the FFT input size $N$ in samples. Therefore, the input size is inversely proportional to RBW:
\begin{equation}
    \mathrm{RBW} = \frac{f_s}{N}
\end{equation}

The ADC card provides \texttt{int16} quantized voltage measurements, with multiple channels interleaved into a single buffer. The raw ADC data is first converted into a proper physical voltage scale using a CUDA kernel, based on the selected hardware voltage range. The result is stored in a single-precision floating-point buffer. Optionally, a windowing function is applied during this type conversion\footnote{We precompute and interleave the distinct windows for each channel, requiring the GPU to merely multiply the windowing buffer with the voltage buffer.}.

For computing FFTs, a \texttt{cuFFT} plan is created at DAQ startup, accounting for the interleaved input format of the multi-channel DAQ. The \texttt{cuFFT} library computes the FFTs for each channel and stores them as de-interleaved continuous data inside a \texttt{float2} buffer. Since the output is complex-valued, it requires twice as much memory per element. The FFT output of a real-valued measurement contains $\lfloor\frac{N}{2}\rfloor + c$ elements, where $N$ is the number of input samples and $c$ is the number of channels.

After computing the FFTs, their absolute values are calculated and accumulated into a separate buffer using a custom CUDA kernel. To track physical quantities such as signal power and its variance, we additionally store the squared and fourth-power magnitude of the FFT spectrum in separate accumulation buffers. The callback function continuously accumulates these powers until a user-specified number of FFTs is reached, after which the buffers are averaged via a division kernel. We store these averaged buffers into an HDF5 file utilizing a multi-threaded file handler. For our specific setup, this file contains the averaged spectrum, the average of the squared spectrum, and the average of the fourth power of the spectrum across two channels.

For an input size of $N$ elements, the minimum required VRAM, excluding internal \texttt{cuFFT} workspace overhead, is given by:
\begin{equation}
\textrm{VRAM} \approx N \Big[\text{sizeof(int16)} + 5 \cdot \text{sizeof(float)}\Big] \approx 22N
\end{equation}
where $\textrm{VRAM}$ is in units of bytes. This calculation accounts for the raw input buffer, the intermediate processing stages, and the complex spectral output, where the $\lfloor\frac{N}{2}\rfloor + c$ complex coefficients are approximately equivalent to $N$ floating point values.

\subsubsection{Triple-GPU Setup}

The single-GPU setup is severely limited in the achievable RBW by the onboard VRAM of the hardware. Most consumer GPUs are not designed for applications with massive contiguous memory requirements. We circumvent this by utilizing multiple GPUs. For our current WISPLC \cite{WISPLC} deployment, we utilize two \SI{11}{\giga\byte} VRAM NVIDIA RTX 2080 TI cards \cite{nvidia_rtx2080ti} and one \SI{16}{\giga\byte} VRAM NVIDIA RTX A4000 card \cite{nvidia_rtxa4000}.

This multi-GPU setup achieves an increase in the available memory at the expense of additional overhead to handle the data transfer across several GPUs. It is not possible to launch kernels or \texttt{cuFFT} computations on buffers spread over different physical GPUs. While it is theoretically simpler to split the workload across GPUs by physical channels to preserve memory efficiency, this would require either prior de-interleaving of the ADC data by a separate host CPU or complex strided memory copy schemes. Instead, we split the workload sequentially into three stages: type conversion, FFT computation, and averaging. These stages are distributed across three GPUs, with the FFT GPU possessing the most VRAM. The disadvantage of this approach is the loss of VRAM efficiency, as it requires mirrored transfer buffers between the successive GPUs. In the absence of a shared PCIe bridge or NVIDIA NVLink interconnect, data transfers are routed through host RAM. This creates a performance bottleneck as discussed in Section \ref{sec:benchmarks}.

Despite this inefficiency, we achieve a target resolution bandwidth of roughly \SI{0.1}{\hertz} with a sample rate of \SI[per-mode=symbol]{62}{\mega\sample\per\second} per channel on a two-channel setup. This equates to an input size of \SI{1}{\giga\byte} of \texttt{int16} data per channel, or a \SI{4}{\giga\byte} total FFT input size. Both the VRAM of the first and second GPU are allocated almost completely under these parameters. However, the final GPU requires significantly less memory. We leverage this excess capacity on the third GPU to compute and store real-time statistics of the acquired data, such as average total power, variances, and phase alignments.

\begin{figure}
    \centering
    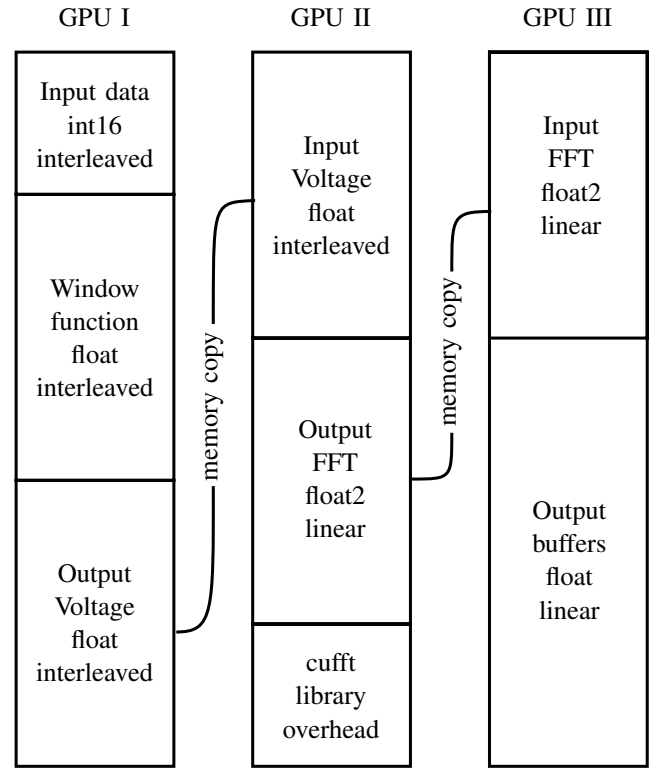
    \caption{Overview of the sequential memory layout and workload distribution across the three GPUs.}
    \label{fig:mem_three}
\end{figure}

\section{Verification}
\label{sec:verification}

To verify the absence of dead time in the data acquisition system \textit{in-situ}, two complementary validation tests were performed. The first test is designed to probe potential discontinuities occurring during file saving and buffer transitions. The second test provides a full-stack validation of the zero-dead-time performance of the acquisition chain. Finally, the long-term stability of the system was assessed through continuous operation of the WISPLC setup over several months, during which no interruptions, data loss, or performance degradation were observed.

\subsection{File Saving and Buffer Boundary Analysis}

\begin{figure}
\label{fig:sine}
\centering
\includegraphics[width=\linewidth]{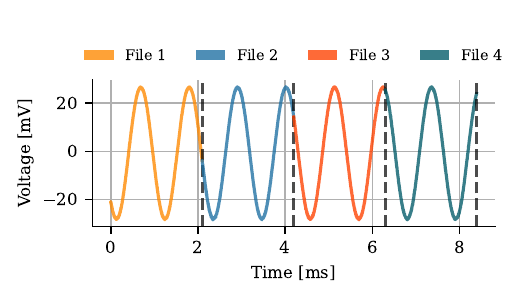}
\caption{Experimental verification of zero-dead-time continuous data acquisition. A continuous \SI{900}{\hertz} sinusoidal signal was injected into the ADC. The plot displays four consecutively saved data files, separated by vertical dashed lines. The perfect phase continuity at the exact sample boundary demonstrates that no data is dropped during hardware buffer wraparounds or software callback executions.}
\label{fig:sine_injection}
\end{figure}

As a first test, continuous sinusoidal signals at various frequencies were injected into the ADC card. These signals were sampled at a rate of \SI[per-mode=symbol]{250}{\mega\sample\per\second} to analyze the phase continuity at the boundary between consecutive data files. In total, six frequencies were evaluated, generating 5000 files per frequency with a file size of \SI{2}{\mega\byte} each. Fig.~\ref{fig:sine_injection} illustrates a continuous sine wave at \SI{900}{\hertz} spanning four distinct files, where the vertical dashed lines denote the exact boundaries between files. This plot confirms the phase continuity for a representative sample. However, to ensure this performance is consistent throughout the entire dataset, a broader statistical verification is required.

\begin{figure*}
\centering
\includegraphics[width=\linewidth]{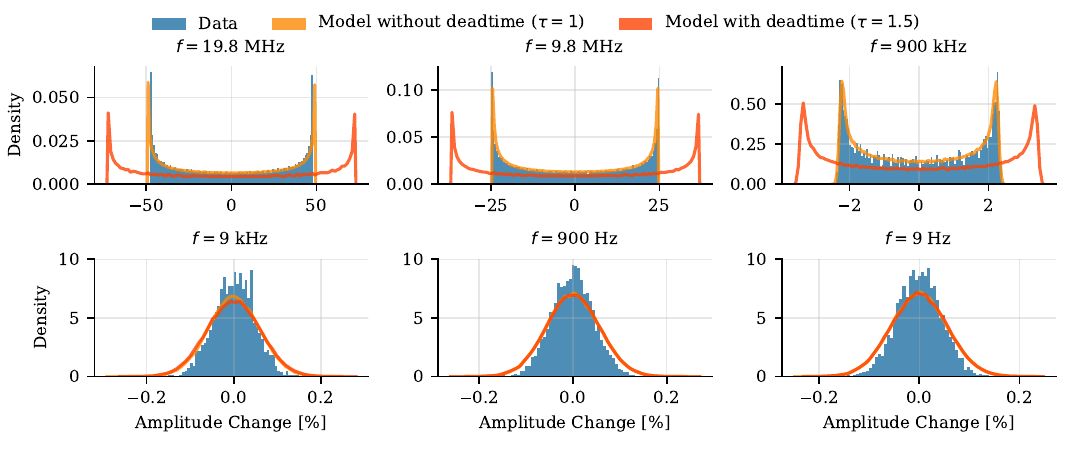}
\caption{Histograms of the amplitude change across consecutive file boundaries for six injected frequencies: \SI{19.8}{\mega\hertz}, \SI{9.8}{\mega\hertz}, \SI{900}{\kilo\hertz}, \SI{9}{\kilo\hertz}, \SI{900}{\hertz}, and \SI{9}{\hertz}. The signals were sampled at \SI[per-mode=symbol]{250}{\mega\sample\per\second}. The experimental data are overlaid with Monte Carlo simulations representing both a strict zero-dead-time condition and a finite dead-time condition with a 0.5 sample delay. The symmetric spread around zero in the top row, together with its perfect alignment with the zero-dead-time simulation, demonstrates continuous acquisition capability. The bottom row shows the expected distributions dominated by ADC quantization and system noise at lower frequencies.}
\label{fig:sine_dist}
\end{figure*}

In order to rigorously quantify the continuity between files, the amplitude difference at every file boundary was computed. This procedure yields amplitude distributions that are sensitive to dead time, which enables the comparison between data and statistical simulations. The resulting sample distributions are presented in Fig.~\ref{fig:sine_dist}. To complement the experimental measurements, a Monte Carlo simulation of the measurement process was developed to model the expected amplitude difference distribution under both zero and finite dead-time conditions. The simulation generates consecutive discrete-time samples of a sinusoidal signal at frequency $f$. The final sample of the $n$th file is defined as
\begin{equation}
x_1(n) = \sin\left(2\pi f\frac{nN}{f_s}\right),
\end{equation}
where $N$ is the length of each file. The subsequent sample, representing the first data point of the next file, is modeled as
\begin{equation}
x_2(n) = \sin\left(2 \pi f \frac{nN + \tau}{f_s}\right) + \eta,
\end{equation}
where $\tau =1$ for the dead-time free model. To model a system with finite dead time, a fractional delay was introduced such that $\tau = 1.5$. Furthermore, Gaussian noise was added to the second signal to match the experimental conditions, where the noise term is modeled as
\begin{equation}
\eta \sim \mathcal{N}(0, \sigma_v^2).
\end{equation}
The noise standard deviation $\sigma_v$ is estimated directly from the experimental data by evaluating the spectral noise floor of the corresponding FFT after the removal of the dominant sinusoidal component. The simulated observable is defined as the amplitude difference across the file boundary,
\begin{equation}
\Delta x = x_2 - x_1,
\end{equation}
which is evaluated over an ensemble of $K = 10^5$ independent realizations. The resulting distribution is then compared directly to the experimentally measured distributions under identical sampling conditions.

The experimental results are in agreement with the zero-dead-time simulations. As depicted in the upper row of Fig.~\ref{fig:sine_dist}, the expected distribution scales appropriately with frequency, and the data adequately tracks the zero-dead-time model. Conversely, the lower row exhibits noise-dominated distributions that lack frequency dependence. Due to a shallow signal slope at low frequencies, the expected voltage change over a single \SI{8}{\nano\second} sample interval is smaller than the system noise.

\subsection{End-to-end Zero-Dead Time Validation}
In the second test, a \SI{3}{\mega\hertz} sinusoidal signal was injected into the ADC while synchronizing both the signal generator and the ADC card to a shared \SI{10}{\mega\hertz} reference clock. This signal was sampled at \SI[per-mode=symbol]{125}{\mega\sample\per\second} for a continuous duration of about 10 minutes\footnote{Unlike frequency-domain spectra, time-domain data cannot be easily averaged to reduce data rates. Consequently, continuous time-domain acquisition requires storage hardware capable of write speeds exceeding the raw data rate. The \SI[per-mode=symbol]{125}{\mega\sample\per\second} rate was chosen to accommodate the maximum sustained write speed of the available solid-state drives, as testing at the full \SI[per-mode=symbol]{500}{\mega\sample\per\second} rate requires a sustained throughput exceeding \SI[per-mode=symbol]{1000}{\mega\byte\per\second} for the entire 10-minute duration.}. Using the first \SI{20}{\mega\byte} of data (corresponding to approximately \SI{0.8}{\second}), the initial phase $\phi_0$ of the \SI{3}{\mega\hertz} signal was estimated from the time-domain samples $x[n]$ using
\begin{equation}
\label{eq:phi}
\phi_0 = \arctan\!\left(
\frac{\left\langle x[n], \sin(2 \pi f \frac{n}{f_s})\right\rangle}
{\left\langle x[n], \cos(2 \pi f \frac{n}{f_s})\right\rangle}
\right),
\end{equation}
where $n$ is the discrete sample index and $\langle \cdot,\cdot \rangle$ denotes the inner product over the available samples.

\begin{figure*}
\centering
\includegraphics[width=\linewidth]{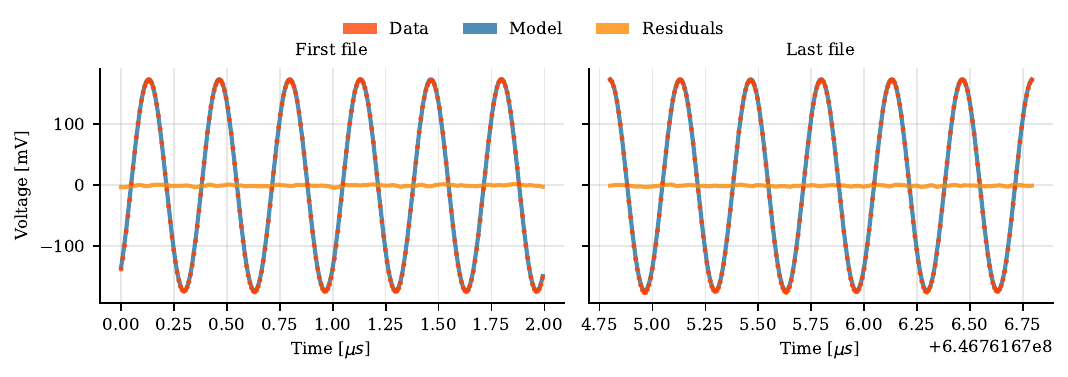}
\caption{Plot of the captured \SI{3}{\mega\hertz} sinusoidal signal superimposed with a mathematical model derived exclusively from the initial \SI{0.8}{\second} of acquired data. The left panel shows the first 250 samples of the initial file, while the right panel shows the last 250 samples of the final file recorded around 10 minutes later. The perfect phase overlap of the extrapolated model and the data in the right panel indicates zero dead time throughout the entire data acquisition run.}
\label{fig:sine_long}
\end{figure*}

Using the initial phase $\phi_0$ and the known frequency $f = \SI{3}{\mega\hertz}$, a continuous sinusoidal model was extrapolated across the entire 10-minute acquisition time. This extrapolated model remains in near-perfect agreement with the recorded data for the full duration. A comparison of the first and final data segments against the model is presented in Fig.~\ref{fig:sine_long}. Due to inherent ADC clock jitter, estimating an absolute phase shift over 10 minutes is challenging. However, it is possible to constrain the data loss by constraining the phase shift. Dead time would manifest as a systematic, cumulative timing discontinuity, whereas ADC clock jitter produces zero-mean phase fluctuations. Using the acquired data, the maximum phase drift is constrained to be below one-hundredth of a phase. This constraint corresponds to a maximum hypothetical dead time of
\begin{equation}
    T_{\mathrm{dead}} = \frac{\Delta\phi}{2\pi f} \leq \frac{1}{100 f} \approx \SI{3.33}{\nano\second},
\end{equation}
which is less than the duration of a single sample interval $t_s$, given by
\begin{equation}
    t_s = \frac{1}{f_s} = \SI{8}{\nano\second}.
\end{equation}
This bounds the fractional data loss $L$ over the total acquisition time $T_{\mathrm{tot}}$ to
\begin{equation}
    L = \frac{T_{\mathrm{dead}}}{T_{\mathrm{tot}}} \leq \frac{\SI{3.33}{\nano\second}}{\SI{6e11}{\nano\second}} \approx 5.6\times 10^{-12}.
\end{equation}
Therefore, it is reasonable to establish an upper bound on the fractional dead time of the acquisition system on the order of $10^{-12}$. Within the resolution limits of the measurement hardware, no evidence of missing samples or timing discontinuities is observed over the full 10-minute acquisition time.

\subsection{Long-Term Stability}
Finally, to validate long-term stability, we deployed the full system, including HDF5 file saving, for extended physics runs during a WISPLC data-taking campaign. The system operated continuously for roughly two months in total, with the longest uninterrupted run lasting one full month. The software functioned flawlessly without any crashes or memory corruption. Additionally, the onboard ADC hardware maintained a highly stable operating temperature, with minimal fluctuations around \SI{48}{\celsius} throughout the entire duration.

\section{Benchmarks}
\label{sec:benchmarks}

To evaluate the performance and reliability of the proposed data acquisition platform, we conducted a series of stress tests designed to probe computational throughput, memory transfer bottlenecks, and long-term stability. By replacing the UI thread with an automated benchmark controller thread, we systematically started and stopped the software across different input parameters. We benchmarked the triple GPU setup by acquiring a fixed data volume and computing the FFT in real time. For each benchmark size, we performed 5 iterations utilizing the maximum hardware sample rate. We ran a total of 100 benchmarks, each with a different input size.

\begin{figure*}
\centering
\includegraphics[width=\linewidth]{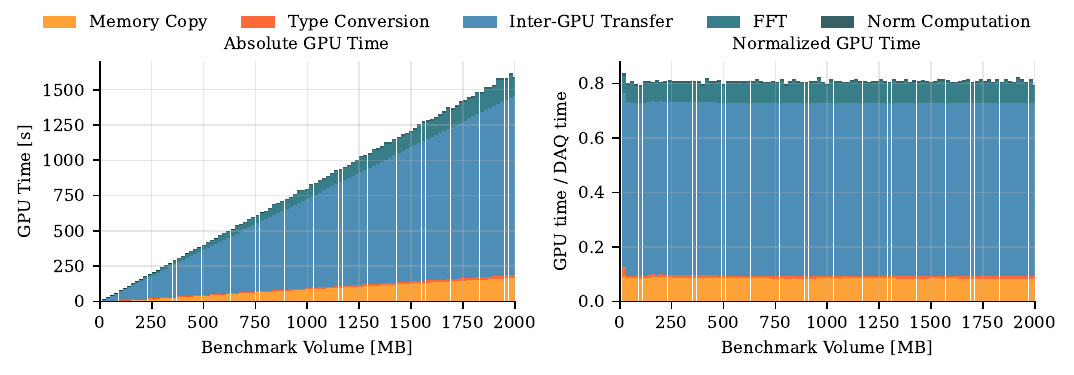}
\caption{Performance benchmarks of the triple GPU setup across 100 different input data volumes. The left panel details the absolute time spent on individual computation and inter-GPU memory transfer stages. The right panel displays the total processing time normalized by the DAQ time at a maximum sample rate of \SI[per-mode=symbol]{500}{\mega\sample\per\second}. The constant fraction demonstrates that the processing scales proportionally with the input data volume, dominated by transfer latencies rather than algorithmic complexity.}
\label{fig:benchmark}
\end{figure*}

The average time spent on each computation step is detailed in Fig.~\ref{fig:benchmark}. We normalized the total GPU computation time by the DAQ time, which is the time required by the hardware to acquire the respective data chunk. We performed this normalization using the maximum sample rate of \SI[per-mode=symbol]{500}{\mega\sample\per\second}, which is equivalent to a data throughput of \SI[per-mode=symbol]{1}{\giga\byte\per\second}. While the fast Fourier transform algorithm scales with $\mathcal{O}(N \log N)$ complexity, the actual computation time is heavily dwarfed by the inter-GPU memory transfer latencies. All other custom processing kernels operate with $\mathcal{O}(N)$ time complexity. Consequently, as shown in the right panel of Fig.~\ref{fig:benchmark}, the entire real-time processing consumes a fixed fraction of the DAQ time regardless of the input size.

At the maximum sample rate, the data processing requires approximately 80\% of the available DAQ time. However, when operating at a lower sample rate, such as the \SI[per-mode=symbol]{124}{\mega\sample\per\second} utilized for the WISPLC experiment, the usage drops to roughly 20\%.

The vast majority of the required processing time is spent on memory transfers. Because the PCIe bridge of our host motherboard does not support direct GPU peer-to-peer memory transfers, we must route the data twice through the host CPU RAM. Utilizing a motherboard with a proper PCIe topology or incorporating NVIDIA NVLink bridges would immediately halve these transfer times. This hardware upgrade would reduce the total DAQ time fraction to approximately 12\% and 50\% for the \SI[per-mode=symbol]{124}{\mega\sample\per\second} and \SI[per-mode=symbol]{500}{\mega\sample\per\second} configurations, respectively.

\section{Conclusion}
\label{sec:conclusion}

The ever-increasing data rates of modern high-throughput physics experiments necessitate a shift from rigid hardware processing to more flexible and scalable software architectures. We have presented a modular data acquisition and real-time processing platform that addresses these challenges by combining high-performance PCIe digitizers with consumer-grade GPUs. By utilizing a highly scalable multi-threaded POSIX (pthreads) architecture, the system efficiently distributes data acquisition and memory transfer loads to maintain a zero-dead-time pipeline for fast digitizers. Furthermore, the software provides a high degree of flexibility through a user-defined callback framework. This architecture facilitates the seamless integration of software-defined logic, ranging from digital filtering and transient signal detection to continuous fast Fourier transforms and statistical averaging, without altering the core pipeline.

The salient features of the system have been empirically validated through the WISPLC dark matter search experiment at the University of Hamburg. During this deployment, the platform maintained a resolution bandwidth of \SI{0.1}{\hertz} at a sampling rate of \SI[per-mode=symbol]{124}{\mega\sample\per\second} with a bandwidth of \SI{62}{\MHz} across two channels. The implementation of real-time statistical averaging successfully reduced the experimental data storage requirements from \SI{21}{\tera\byte} per day to less than \SI{20}{\tera\byte} per month. Furthermore, our custom Faraday cage and cooling solution ensured high thermal and electromagnetic stability over a continuous month-long acquisition period, with negligible temperature fluctuations and a reduction in environmental noise spurs.

This architecture demonstrates that high-performance GPU-based platforms can effectively replace specialized hardware pipelines like FPGAs for medium to large-scale laboratory setups. The transition to software-defined processing significantly lowers the entry barrier for high-throughput experiments by utilizing standard consumer electronics and established programming frameworks. This flexibility allows for rapid iteration and adaptation to different experimental goals, such as transient signal detection or high-resolution spectral analysis, without requiring hardware redesigns.

\section{Outlook and Perspectives}
\label{sec:outlook}

While the current deployment has successfully demonstrated the viability of software-defined data acquisition for specific physics experiments, the underlying architecture offers opportunities for a broad range of applications. In conjunction with appropriate RF equipment, this platform can already replace multiple types of standard laboratory instruments, such as spectrum analyzers and time-domain data acquisition systems. This significantly simplifies the data acquisition and pre-processing stages. Real-time averaging and the computation of basic derived quantities drastically decrease data storage requirements and, consequently, total experiment costs.

We plan to implement additional real-time processing options, including matched filtering for an upcoming high-frequency gravitational wave detector at the University of Hamburg. To improve usability for researchers unfamiliar with C and CUDA programming, we are planning to develop a visual interface for easier customization of the data processing pipeline. Furthermore, we intend to integrate a network library to decouple the software into a client-server architecture, removing the user interface computing load from the host server while retaining real-time control.

Finally, the software was deliberately designed to accommodate future hardware expansions. By integrating a digital-to-analog converter (DAC) card, an external mixing unit, directional couplers, and GPIO-controlled SMA switches, the system could perform custom signal synthesis and scatter-parameter network analysis. The inclusion of heterodyne downmixing further enables the platform to probe signals in the \si{\giga\hertz} range, effectively extending the experimental frequency reach far beyond the hardware Nyquist limit of the digitizer. This extended high-frequency architecture will be implemented in the ADAMOS project to facilitate axion dark matter searches at \SI{20}{\giga\hertz} \cite{Maroudas_2026_adamos}. Considering the cost-effectiveness and flexibility of this platform, it has the potential to replace expensive, rigid laboratory devices while retaining full raw data acquisition capabilities.

\section*{Acknowledgments}
M.M. acknowledges funding by the Deutsche Forschungsgemeinschaft (DFG, German Research Foundation) under Germany’s Excellence Strategy – EXC 2121 ``Quantum Universe" – 390833306. We thank Marko Ekmedžić for designing the ADC enclosure.

\bibliographystyle{IEEEtran}

\bibliography{ref}

\end{document}

%% file: thread_structure.pdf_tex
\begingroup%
  \makeatletter%
  \providecommand\color[2][]{%
    \errmessage{(Inkscape) Color is used for the text in Inkscape, but the package 'color.sty' is not loaded}%
    \renewcommand\color[2][]{}%
  }%
  \providecommand\transparent[1]{%
    \errmessage{(Inkscape) Transparency is used (non-zero) for the text in Inkscape, but the package 'transparent.sty' is not loaded}%
    \renewcommand\transparent[1]{}%
  }%
  \providecommand\rotatebox[2]{#2}%
  \newcommand*\fsize{\dimexpr\f@size pt\relax}%
  \newcommand*\lineheight[1]{\fontsize{\fsize}{#1\fsize}\selectfont}%
  \ifx\svgwidth\undefined%
    \setlength{\unitlength}{236.59085732bp}%
    \ifx\svgscale\undefined%
      \relax%
    \else%
      \setlength{\unitlength}{\unitlength * \real{\svgscale}}%
    \fi%
  \else%
    \setlength{\unitlength}{\svgwidth}%
  \fi%
  \global\let\svgwidth\undefined%
  \global\let\svgscale\undefined%
  \makeatother%
  \begin{picture}(1,0.97562998)%
    \lineheight{1}%
    \setlength\tabcolsep{0pt}%
    \put(0,0){\includegraphics[width=\unitlength,page=1]{thread_structure.pdf}}%
    \put(0.11856613,0.91879941){\makebox(0,0)[t]{\lineheight{1.25}\smash{\begin{tabular}[t]{c}UI Thread\end{tabular}}}}%
    \put(0,0){\includegraphics[width=\unitlength,page=2]{thread_structure.pdf}}%
    \put(0.11856613,0.0315017){\makebox(0,0)[t]{\lineheight{1.25}\smash{\begin{tabular}[t]{c}UI Thread\end{tabular}}}}%
    \put(0,0){\includegraphics[width=\unitlength,page=3]{thread_structure.pdf}}%
    \put(0.40704965,0.92190398){\makebox(0,0)[t]{\lineheight{1.25}\smash{\begin{tabular}[t]{c}DAQ Thread\end{tabular}}}}%
    \put(0,0){\includegraphics[width=\unitlength,page=4]{thread_structure.pdf}}%
    \put(0.40757851,0.03460627){\makebox(0,0)[t]{\lineheight{1.25}\smash{\begin{tabular}[t]{c}DAQ Thread\end{tabular}}}}%
    \put(0,0){\includegraphics[width=\unitlength,page=5]{thread_structure.pdf}}%
    \put(0.80515222,0.91879945){\makebox(0,0)[t]{\lineheight{1.25}\smash{\begin{tabular}[t]{c}ADC Thread\end{tabular}}}}%
    \put(0,0){\includegraphics[width=\unitlength,page=6]{thread_structure.pdf}}%
    \put(0.80515222,0.0315017){\makebox(0,0)[t]{\lineheight{1.25}\smash{\begin{tabular}[t]{c}ADC Thread\end{tabular}}}}%
    \put(0,0){\includegraphics[width=\unitlength,page=7]{thread_structure.pdf}}%
    \put(0.25411007,0.81038271){\makebox(0,0)[t]{\lineheight{1.25}\smash{\begin{tabular}[t]{c}START\end{tabular}}}}%
    \put(0,0){\includegraphics[width=\unitlength,page=8]{thread_structure.pdf}}%
    \put(0.59976744,0.72682472){\makebox(0,0)[t]{\lineheight{1.25}\smash{\begin{tabular}[t]{c}pthread\_create()\end{tabular}}}}%
    \put(0,0){\includegraphics[width=\unitlength,page=9]{thread_structure.pdf}}%
    \put(0.91357172,0.65692893){\makebox(0,0)[t]{\lineheight{1.25}\smash{\begin{tabular}[t]{c}callback()\end{tabular}}}}%
    \put(0,0){\includegraphics[width=\unitlength,page=10]{thread_structure.pdf}}%
    \put(0.25538082,0.52242224){\makebox(0,0)[t]{\lineheight{1.25}\smash{\begin{tabular}[t]{c}plot()\end{tabular}}}}%
    \put(0,0){\includegraphics[width=\unitlength,page=11]{thread_structure.pdf}}%
    \put(0.91357172,0.45524087){\makebox(0,0)[t]{\lineheight{1.25}\smash{\begin{tabular}[t]{c}callback()\end{tabular}}}}%
    \put(0,0){\includegraphics[width=\unitlength,page=12]{thread_structure.pdf}}%
    \put(0.25538082,0.31801987){\makebox(0,0)[t]{\lineheight{1.25}\smash{\begin{tabular}[t]{c}plot()\end{tabular}}}}%
    \put(0,0){\includegraphics[width=\unitlength,page=13]{thread_structure.pdf}}%
    \put(0.25403425,0.23446176){\makebox(0,0)[t]{\lineheight{1.25}\smash{\begin{tabular}[t]{c}STOP\end{tabular}}}}%
    \put(0,0){\includegraphics[width=\unitlength,page=14]{thread_structure.pdf}}%
    \put(0.59733667,0.15090355){\makebox(0,0)[t]{\lineheight{1.25}\smash{\begin{tabular}[t]{c}STOP\end{tabular}}}}%
  \end{picture}%
\endgroup%

%% file: GPU_THREE_MEMLAYOUT.pdf_tex
\begingroup%
  \makeatletter%
  \providecommand\color[2][]{%
    \errmessage{(Inkscape) Color is used for the text in Inkscape, but the package 'color.sty' is not loaded}%
    \renewcommand\color[2][]{}%
  }%
  \providecommand\transparent[1]{%
    \errmessage{(Inkscape) Transparency is used (non-zero) for the text in Inkscape, but the package 'transparent.sty' is not loaded}%
    \renewcommand\transparent[1]{}%
  }%
  \providecommand\rotatebox[2]{#2}%
  \newcommand*\fsize{\dimexpr\f@size pt\relax}%
  \newcommand*\lineheight[1]{\fontsize{\fsize}{#1\fsize}\selectfont}%
  \ifx\svgwidth\undefined%
    \setlength{\unitlength}{237.80091053bp}%
    \ifx\svgscale\undefined%
      \relax%
    \else%
      \setlength{\unitlength}{\unitlength * \real{\svgscale}}%
    \fi%
  \else%
    \setlength{\unitlength}{\svgwidth}%
  \fi%
  \global\let\svgwidth\undefined%
  \global\let\svgscale\undefined%
  \makeatother%
  \begin{picture}(1,1.19899106)%
    \lineheight{1}%
    \setlength\tabcolsep{0pt}%
    \put(0.12665646,1.05260665){\color[rgb]{0,0,0}\makebox(0,0)[t]{\lineheight{1.25}\smash{\begin{tabular}[t]{c}Input data\\int16\\interleaved\end{tabular}}}}%
    \put(0.12665652,0.74319245){\color[rgb]{0,0,0}\makebox(0,0)[t]{\lineheight{1.25}\smash{\begin{tabular}[t]{c}Window\\function\\float\\interleaved\end{tabular}}}}%
    \put(0.12627173,0.29367475){\color[rgb]{0,0,0}\makebox(0,0)[t]{\lineheight{1.25}\smash{\begin{tabular}[t]{c}Output\\Voltage \\float\\interleaved \end{tabular}}}}%
    \put(0.4997131,0.9677284){\color[rgb]{0,0,0}\makebox(0,0)[t]{\lineheight{1.25}\smash{\begin{tabular}[t]{c}Input \\Voltage\\float\\interleaved\end{tabular}}}}%
    \put(0.50572379,0.5206802){\color[rgb]{0,0,0}\makebox(0,0)[t]{\lineheight{1.25}\smash{\begin{tabular}[t]{c}Output \\FFT\\float2\\linear\end{tabular}}}}%
    \put(0.49902929,0.15245685){\color[rgb]{0,0,0}\makebox(0,0)[t]{\lineheight{1.25}\smash{\begin{tabular}[t]{c}cufft\\library \\overhead\end{tabular}}}}%
    \put(0.87759782,0.99913473){\color[rgb]{0,0,0}\makebox(0,0)[t]{\lineheight{1.25}\smash{\begin{tabular}[t]{c}Input \\FFT\\float2\\linear\end{tabular}}}}%
    \put(0.8748613,0.39240367){\color[rgb]{0,0,0}\makebox(0,0)[t]{\lineheight{1.25}\smash{\begin{tabular}[t]{c}Output \\buffers\\float\\linear\end{tabular}}}}%
    \put(0,0){\includegraphics[width=\unitlength,page=1]{GPU_THREE_MEMLAYOUT.pdf}}%
    \put(0.31983288,0.57061588){\color[rgb]{0,0,0}\rotatebox{90}{\makebox(0,0)[t]{\lineheight{1.25}\smash{\begin{tabular}[t]{c}memory copy\\\end{tabular}}}}}%
    \put(0,0){\includegraphics[width=\unitlength,page=2]{GPU_THREE_MEMLAYOUT.pdf}}%
    \put(0.69597574,0.6564067){\color[rgb]{0,0,0}\rotatebox{90}{\makebox(0,0)[t]{\lineheight{1.25}\smash{\begin{tabular}[t]{c}memory copy\end{tabular}}}}}%
    \put(0,0){\includegraphics[width=\unitlength,page=3]{GPU_THREE_MEMLAYOUT.pdf}}%
    \put(0.12588657,1.16925733){\color[rgb]{0,0,0}\makebox(0,0)[t]{\lineheight{1.25}\smash{\begin{tabular}[t]{c}GPU I\\\end{tabular}}}}%
    \put(0.49894313,1.16925733){\color[rgb]{0,0,0}\makebox(0,0)[t]{\lineheight{1.25}\smash{\begin{tabular}[t]{c}GPU II\\\end{tabular}}}}%
    \put(0.87235994,1.16925717){\color[rgb]{0,0,0}\makebox(0,0)[t]{\lineheight{1.25}\smash{\begin{tabular}[t]{c}GPU III\\\end{tabular}}}}%
  \end{picture}%
\endgroup%